\documentclass[aps,pre,twocolumn,superscriptaddress,10pt]{revtex4-1} 
\pdfoutput=1
\usepackage[pdftex]{graphicx} 
\usepackage[breaklinks]{hyperref}
\hypersetup{pdftitle={Adaptive Thouless--Anderson--Palmer equation for higher-order Markov random fields},colorlinks,citecolor=[rgb]{.547,.129,.332},urlcolor=[rgb]{.094,.223,.617},linkcolor=[rgb]{.547,.129,.332}}

\usepackage{amsmath,newtxtext,newtxmath,bm}

\newcommand{\sref}[1]{Section~\ref{#1}}
\newcommand{\eref}[1]{Eq.~\eqref{#1}}

\newcommand{\aref}[1]{Appendix~\ref{#1}}

\begin{document}

\title{Adaptive Thouless--Anderson--Palmer equation for higher-order Markov random fields}

\author{Chako Takahashi}
\email{chako@dc.tohoku.ac.jp}
\affiliation{Graduate School of Information Sciences, Tohoku University, 6-3-09 Aoba, Aramaki, Aoba, Sendai, Miyagi 980-8579, Japan}

\author{Muneki Yasuda}
\affiliation{Graduate School of Science and Engineering, Yamagata University, 4-3-16 Jonan, Yonezawa, Yamagata 992-8510, Japan}

\author{Kazuyuki Tanaka}
\affiliation{Graduate School of Information Sciences, Tohoku University, 6-3-09 Aoba, Aramaki, Aoba, Sendai, Miyagi 980-8579, Japan}

\begin{abstract}
 The adaptive Thouless--Anderson--Palmer (TAP) mean-field approximation is one of the advanced mean-field approaches, and it is known as a powerful accurate method for Markov random fields (MRFs) with quadratic interactions (pairwise MRFs).
 In this study, an extension of the adaptive TAP approximation for MRFs with many-body interactions (higher-order MRFs) is developed.
 We show that the adaptive TAP equation for pairwise MRFs is derived by naive mean-field approximation with diagonal consistency.
 Based on the equivalence of the approximate equation obtained from the naive mean-field approximation with diagonal consistency and the adaptive TAP equation in pairwise MRFs, we formulate approximate equations for higher-order Boltzmann machines, which is one of simplest higher-order MRFs, via the naive mean-field approximation with diagonal consistency.
\end{abstract}

\maketitle

\section{Introduction}
\label{sec:introduction}

A Markov random field (MRF) is known as an important probabilistic graphical model in various scientific fields.
There are a large variety of applications of MRFs, for example, in computer vision~\cite{MRFLi1995, MRFBlake2011}, engineering~\cite{frey1998graphical}, machine learning~\cite{wainwright2008graphical,koller2009probabilistic}, information sciences~\cite{mezard2009information,nishimori2001}, and statistical physics.
A Boltzmann machine~\cite{ackley1985learning}, which is a kind of an MRF, is known as the fundamental probabilistic model in such fields.
A typical Boltzmann machine is the same as an Ising model in statistical physics.
A Boltzmann machine defined on a complete bipartite graph called a restricted Boltzmann machine (RBM) has frequently been used in deep learning~\cite{hinton2006reducing,salakhutdinov2010efficient}.
Statistical operations, such as the computation of expectations in MRFs, are computationally intractable in most cases because they require summations over all possible states of variables.
Hence, we use approximate techniques, such as the Markov chain Monte Carlo (MCMC) method, for statistical computations.
For RBMs, approximate learning methods based on MCMC sampling, such as contrastive divergence~\cite{Hinton2002}, have been proposed and successfully employed.
They alleviate the computational intractability by using conditional independence of RBMs.

Mean-field approximations are effective for MRFs~\cite{meanfield}.
Various mean-field-based methods have been developed in statistical mechanics, for example, the naive mean-field approximation, Thouless--Anderson--Palmer (TAP) approximation~\cite{Morita1976,thouless1977solution}, Bethe approximation (or loopy belief propagation)~\cite{Bethe1935,GBP2001}, and the adaptive TAP approximation~\cite{adaTAP1,adaTAP2}.
Such mean-field methods allow for obtaining the approximate expectations of random variables in MRFs.
In particular, the adaptive TAP approximation is known as one of the most powerful accurate methods in dense systems.
The aim of this study is to extend the adaptive TAP approximation.
Here and hereinafter, the term ``adaptive TAP approximation'' indicates the approach by Opper and Winther~\cite{adaTAP1,adaTAP2}.

The linear response relation~\cite{KR1998} is an important technique for obtaining the accurate approximations of higher-order expectations.
We can calculate such approximations from the expectations obtained by the mean-field methods using the linear response relation.
For instance, susceptibilities (or covariances) are obtained using local magnetizations (or one-variable expectations) by utilizing the linear response relation.
A message-passing type of algorithm based on the linear response relation is known as susceptibility propagation (SusP)~\cite{mezard2009constraint} in statistical physics and as variational linear response in machine learning~\cite{WellingTeh2004}.
However, algorithms that use linear response relation simply such as SusP experience \textit{the diagonal inconsistency problem}~\cite{KR1998,tanaka1998mean,I-SusP}.
In an Ising model, the second-order moment of variable $\langle x_i^2 \rangle$ should be unity because variable $x_i$ takes values of $-1$ or $+1$.
However, the second-order moment obtained by such algorithms is not unity.
Improved SusP (I-SusP), which is an improved version of SusP, was proposed by two of the authors to solve this problem in the context of SusP~\cite{I-SusP,Yasuda2013,Yasuda2014}.
I-SusP allows for using the linear response relation while maintaining diagonal consistency.
This improves the approximation accuracy.
Similar to SusP, I-SusP can be combined with various mean-field methods such as the ones listed above.

The demand for higher-order MRFs (MRFs with higher-order interactions) is growing continuously, particularly in computer vision~\cite{wang2013markov,Huang_2015_ICCV}.
However, we cannot straightforwardly apply the adaptive TAP approximation to higher-order MRFs
because in the conventional approaches to the adaptive TAP approximation, the energy function has to be written in a quadratic form with respect to the variables.
It was found that the results obtained by the adaptive TAP approximation and I-SusP with the naive mean-field approximation are the same in an Ising model~\cite{I-SusP,jack2013meanfield}.
Based on this, we expect the adaptive TAP approximation and I-SusP to be equivalent in other models.
If this prediction is justified, we can construct the adaptive TAP approximation via the same calculation as I-SusP for any case.
This implies that we can construct the adaptive TAP approximation for higher-order MRFs because I-SusP can be applied to various models, including higher-order MRFs.
Here and hereinafter, the term ``I-SusP'' indicates the extended message-passing algorithm of belief propagation proposed by two of the authors~\cite{I-SusP}.

The goal of this study is to formulate the adaptive TAP approximation for higher-order MRFs.
In order to achieve this, first, we show the equivalence of the adaptive TAP approximation and the naive mean-field approximation with diagonal consistency in MRFs with quadratic energy functions (i.e., MRFs with Ising or non-Ising variables).
This shows that the equivalence is justified at least in the models to which the adaptive TAP approximation can be straightforwardly applied.
This fact strongly supports our prediction about the equivalence, based on which we tentatively accept our prediction as the ansatz.
After that, we formulate the adaptive-TAP-like equation for higher-order Boltzmann machines, which is one of simplest higher-order MRFs, via the naive mean-field approximation with diagonal consistency.
As the equivalence of the adaptive TAP approximation and the naive mean-field approximation with diagonal consistency has not been rigorously proven yet, we use the word ``like'' here and hereinafter.
The term ``naive mean-field approximation with diagonal consistency'' indicates the approach employed in this study, which follows the same computational procedure as I-SusP, however, it is conceptually different from I-SusP.

The remainder of this paper is organized as follows:
In \sref{sec:pairwiseMRF}, we consider a pairwise MRF with a quadratic energy function.
We introduce Gibbs free energy (GFE), which is a dual representation of Helmholtz free energy, for the pairwise MRF in \sref{sec:GibbsFreeEnergy}.
We derive the adaptive TAP free energy for the pairwise MRF using the GFE presented in \sref{sec:AdaptiveTAPEquation_pairwiseMRF}.
In \sref{sec:I-SusP_pairwiseMRF}, we derive the naive mean-field approximation with diagonal consistency for the pairwise MRF
and subsequently show the equivalence of the approximate equation given by this approach and the adaptive TAP equation derived in \sref{sec:AdaptiveTAPEquation_pairwiseMRF}.
In \sref{sec:adaptiveTAPApproach_higher-order}, we consider a higher-order Boltzmann machine and derive its adaptive-TAP-like equation via the naive mean-field approximation with diagonal consistency (\sref{sec:adaptiveTAPEquation_higher-order}).
In \sref{sec:experiments}, we show through numerical experiments that the naive mean-field approximation with diagonal consistency outperforms the simple naive mean-field approximation in the higher-order MRF, as expected.
Finally, we summarize the study in \sref{sec:conclusion}.

\section{Markov random field\\with quadratic energy function}
\label{sec:pairwiseMRF}

In this section, we consider a pairwise MRF with a quadratic energy function.
Let us consider an undirected graph, $\mathcal{G}(V, E)$, with $n$ vertices,
where $V = \{1, 2, \cdots, n\}$ is the set of all vertices and $E = \{\{i, j\}\}$ is the family of all undirected edges, $\{i,j\}$, in the graph.
Random variables $\bm{x} = \{x_i \in \mathcal{X} \mid i\in V\}$ are assigned to the vertices.
Here, $\mathcal{X}$ is a subset of $\mathbb{R}$.
We define the energy function (or the Hamiltonian) on the graph as
\begin{align}
 H(\bm{x}) := - \sum_{i\in V}h_i x_i + \frac{1}{2}\sum_{i\in V}d_i x_i^2 - \sum_{\{i, j\}\in E} J_{ij} x_i x_j,
 \label{eq:energy}
\end{align}
where $\bm{h} = \{h_i \mid i\in V\}$ and $\bm{d} = \{d_i \mid i \in V\}$ are bias parameters (or the external fields) and anisotropic parameters, respectively, and $\bm{J} = \{J_{ij} \mid \{i, j\} \in E\}$ are the coupling weight parameters between vertices $i$ and $j$.
We assume that there are no self-interactions ($J_{ii} = 0$, $\forall i \in V$).
All couplings are assumed to be symmetric ($J_{ij} = J_{ji}$).
Throughout this paper, we omit the explicit descriptions of the dependency on $\bm{h}$, $\bm{d}$, and $\bm{J}$.
However, it should be noted that almost all quantities described here and in the following sections depend on model parameters.
Along with \eref{eq:energy}, a pairwise MRF is expressed as
\begin{align}
 P(\bm{x}) := \frac{1}{Z}\exp\big(-H(\bm{x})\big),
 \label{eq:pairwiseMRF}
\end{align}
where
\begin{align}
 Z := \sum_{\bm{x} \in \mathcal{X}^n}\exp\big(-H(\bm{x})\big)
 \label{eq:Z}
\end{align}
is the partition function,
and the summation implies $\sum_{\bm{x} \in \mathcal{X}^n} = \sum_{x_1 \in \mathcal{X}}\sum_{x_2 \in \mathcal{X}} \cdots \sum_{x_n \in \mathcal{X}}$.
When $\mathcal{X}$ is a continuous space, the summation over $\bm{x}$ is replaced by integration.
The inverse temperature is set to one throughout this paper.
We refer to the MRF in \eref{eq:pairwiseMRF} as the quadratic MRF.
Note that \eref{eq:pairwiseMRF} becomes the Gaussian MRF (or the Gaussian graphical model)~\cite{GMRF} when $\mathcal{X} = (-\infty, +\infty)$, $d_i > 0$ and the inverse covariance matrix is positive definite.
The $ij$th elements of the covariance matrix are defined by
\begin{align*}
 \big[\bm{C}\big]_{ij} := \left\{
 \begin{array}{ll}
 d_i & (i = j) \\
 - J_{ij} & \left((i,j)\in E\right)\\
 0 & (\mathrm{otherwise})
 \end{array}. \right.
\end{align*}

The Helmholtz free energy of \eref{eq:pairwiseMRF} is expressed as
\begin{align}
F:= - \ln Z.
\label{eq:HelmholtzFreeEnergy}
\end{align}
In the following sections, we introduce a GFE of \eref{eq:pairwiseMRF}, which is a dual representation of $F$.
Moreover, we derive the adaptive TAP equation for \eref{eq:pairwiseMRF} using the GFE.

\subsection{Gibbs free energy and Plefka expansion}
\label{sec:GibbsFreeEnergy}

In this section, we introduce a GFE of the MRF in \eref{eq:pairwiseMRF}.
Let us consider the Kullback--Leibler divergence (KLD) between a test distribution, $Q(\bm{x})$, and the pairwise MRF in \eref{eq:pairwiseMRF}
\begin{align}
 \mathrm{KL}[Q\parallel P] := \sum_{\bm{x} \in \mathcal{X}^n}Q(\bm{x})\ln\frac{Q(\bm{x})}{P(\bm{x})}.
 \label{eq:KLD}
\end{align}
The mean-field approximation can be formulated through the minimization of the KLD~\cite{bilbro1991mean}.
The minimization of the KLD in \eref{eq:KLD} with respect to $Q(\bm{x})$ is equivalent to the minimization of the variational free energy defined by
\begin{align}
 \mathcal{F}[Q] := \sum_{\bm{x} \in \mathcal{X}^n} Q(\bm{x})\ln Q(\bm{x}) + \sum_{\bm{x} \in \mathcal{X}^n}Q(\bm{x})H(\bm{x}),
 \label{eq:variationalFE}
\end{align}
because $\mathrm{KL}[Q\parallel P] = \mathcal{F}[Q] - F$.
By minimizing the variational free energy under the normalization constraint
\begin{align}
\sum_{\bm{x}\in \mathcal{X}^n} Q(\bm{x}) = 1
\label{eq:GibbsConstraints_Normalization}
\end{align}
and moment constraints
\begin{align}
 m_i = \sum_{\bm{x}\in \mathcal{X}^n}x_i Q(\bm{x}), \quad
 v_i = \sum_{\bm{x}\in \mathcal{X}^n}x_i^2 Q(\bm{x}),\quad \forall i \in V,
\label{eq:GibbsConstraints_Moments}
\end{align}
the GFE is obtained as
\begin{align}
G(\bm{m}, \bm{v}) := \min_{Q} \mathcal{F}[Q] \>\> \text{s.t. constraints in Eqs. (\ref{eq:GibbsConstraints_Normalization}) and (\ref{eq:GibbsConstraints_Moments})}.
\label{eq:GibbsFreeEnergy_Original}
\end{align}
The minimum of the GFE with respect to $\bm{m}$ and $\bm{v}$ is equivalent to the Helmholtz free energy, $F = \min_{\bm{m},\bm{v}}G(\bm{m}, \bm{v})$.
Moreover, the $m_i$ and $v_i$ that minimize the GFE coincide with $\langle x_i \rangle$ and $\langle x_i^2 \rangle$, respectively,
where $\langle f(\bm{x}) \rangle := \sum_{\bm{x}\in \mathcal{X}^n}f(\bm{x}) P(\bm{x})$ denotes the exact expectation for the distribution in \eref{eq:pairwiseMRF}.

For the Plefka expansion~\cite{Plefka,expansion} described below, we introduce the auxiliary parameter $\alpha \in [0,1]$ into the energy function in \eref{eq:energy}
as
\begin{align*}
H_{\alpha}(\bm{x}) := -\sum_{i\in V}h_i x_i + \frac{1}{2}\sum_{i\in V}d_i x_i^2 - \alpha\sum_{\{i,j\}\in E}J_{ij}x_i x_j.
\end{align*}
The auxiliary parameter adjusts the effect of the interaction term.
When $\alpha = 1$, $H_{\alpha}(\bm{x})$ is equivalent to $H(\bm{x})$.
We denote the GFE corresponding to $H_{\alpha}(\bm{x})$ by $G_{\alpha}(\bm{m}, \bm{v})$.
By utilizing Lagrange multipliers, $G_{\alpha}(\bm{m}, \bm{v})$ is expressed as
\begin{align}
 &G_{\alpha}(\bm{m}, \bm{v}) = - \sum_{i\in V} h_i m_i + \frac{1}{2} \sum_{i\in V} d_i v_i
 + \max_{\bm{b},\bm{c}}\Big\{ \sum_{i\in V} b_i m_i \nonumber\\
 & - \frac{1}{2}\sum_{i\in V} c_i v_i - \ln \sum_{\bm{x} \in \mathcal{X}^n} \exp \Big( \sum_{i\in V}b_i x_i - \frac{1}{2}\sum_{i\in V}c_i x_i^2 \nonumber\\
 & + \alpha\sum_{\{i,j\}\in E}J_{ij}x_i x_j \Big) \Big\}.
 \label{eq:GibbsFreeEnergy_alpha}
\end{align}
Parameters $\bm{b} = \{b_i \mid i\in V\}$ and $\bm{c} = \{c_i \mid i\in V\}$ originate from the Lagrange multipliers
corresponding to the first and second constraints in \eref{eq:GibbsConstraints_Moments}, respectively.
It is noteworthy that $G_{\alpha}(\bm{m}, \bm{v})$ is equivalent to $G(\bm{m}, \bm{v})$ when $\alpha = 1$.
The maximum conditions for $\bm{b}$ and $\bm{c}$ in \eref{eq:GibbsFreeEnergy_alpha} are obtained as
\begin{align}
 m_i &= \sum_{x_i \in \mathcal{X}}x_i Q_{\alpha}(\bm{x} \mid \bm{b}, \bm{c})
 \label{eq:maximumCondition_m}
\end{align}
and
\begin{align}
 v_i &= \sum_{x_i \in \mathcal{X}}x_i^2 Q_{\alpha}(\bm{x} \mid \bm{b}, \bm{c}),
 \label{eq:maximumCondition_v}
\end{align}
respectively, where
\begin{align*}
Q_{\alpha}(\bm{x} \mid \bm{b}, \bm{c})&:=\frac{1}{Z_{\alpha}(\bm{b}, \bm{c})}
\exp\Big(\sum_{i\in V}b_i x_i - \frac{1}{2}\sum_{i\in V}c_i x_i^2 \nonumber\\
&\quad\;+ \alpha\sum_{\{i,j\}\in E}J_{ij}x_i x_j\Big)
\end{align*}
and $Z_{\alpha}(\bm{b}, \bm{c})$ is the partition function defined in a manner similar to \eref{eq:Z}.
We denote the solutions to Eqs.~(\ref{eq:maximumCondition_m}) and (\ref{eq:maximumCondition_v}) by
$\hat{\bm{m}}(\alpha)$ and $\hat{\bm{v}}(\alpha)$, respectively.
Even though the solutions depend on all parameters in the model,
we omit the description of the dependency, except for $\alpha$, for the convenience of the subsequent analysis.

The Plefka expansion is a perturbative expansion of \eref{eq:GibbsFreeEnergy_alpha} around $\alpha = 0$.
After a perturbative approximation, the corresponding approximation for the original GFE in \eref{eq:GibbsFreeEnergy_Original} is obtained by setting $\alpha = 1$.
Several mean-field approximations are derived based on the Plefka expansion.
For example, the naive mean-field approximation of \eref{eq:GibbsFreeEnergy_Original},
which is referred to as naive mean-field free energy, is obtained as follows:
The expansion up to the first order of \eref{eq:GibbsFreeEnergy_alpha} is
$G_{\alpha}(\bm{m}, \bm{v}) = G_{0}(\bm{m}, \bm{v}) - \alpha \sum_{\{i, j\}\in E}J_{ij} m_i m_j + O(\alpha^2)$.
By setting $\alpha = 1$ in this expanded form, the naive mean-field free energy $G_{\mathrm{naive}}(\bm{m},\bm{v})$ is obtained as
\begin{align}
&G_{\mathrm{naive}}(\bm{m},\bm{v}):= - \sum_{i\in V}h_i m_i + \frac{1}{2}\sum_{i\in V}d_i v_i + \sum_{i\in V} \hat{b}_i(0)m_i\nonumber\\
 & - \frac{1}{2}\sum_{i\in V}\hat{c}_i(0) v_i - \ln Z_0(\hat{\bm{b}}(0), \hat{\bm{c}}(0))
 - \sum_{\{i, j\}\in E}J_{ij} m_i m_j,
\label{eq:NaiveMeanFieldFreeEnergy}
\end{align}
where
\begin{align}
Z_0(\hat{\bm{b}}(0), \hat{\bm{c}}(0))= \prod_{i \in V}\sum_{x_i \in \mathcal{X}}
\exp\Big( \hat{b}_i(0) x_i - \frac{1}{2} \hat{c}_i(0) x_i^2 \Big).
\label{eq:Z_alpha=0}
\end{align}
Based on Eqs.~(\ref{eq:maximumCondition_m}) and (\ref{eq:maximumCondition_v}), $\hat{\bm{b}}(0)$ and $\hat{\bm{c}}(0)$ satisfy
\begin{align}
m_i &= \frac{\sum_{x_i \in \mathcal{X}} x_i \exp\big(\hat{b}_i(0) x_i - \hat{c}_i(0) x_i^2 / 2\big)}
{\sum_{x \in \mathcal{X}} \exp\big(\hat{b}_i(0) x - \hat{c}_i(0) x^2 / 2\big)},
\label{eq:maximumCondition_m_alpha=0}\\
v_i &= \frac{\sum_{x_i \in \mathcal{X}} x_i^2 \exp\big(\hat{b}_i(0) x_i - \hat{c}_i(0) x_i^2 / 2\big)}
{\sum_{x \in \mathcal{X}} \exp\big(\hat{b}_i(0) x - \hat{c}_i(0) x^2 / 2\big)},
\label{eq:maximumCondition_v_alpha=0}
\end{align}
for any $\bm{m}$ and $\bm{v}$.
The naive mean-field equation is obtained from the minimum condition of \eref{eq:NaiveMeanFieldFreeEnergy} with respect to $\bm{m}$ and $\bm{v}$.
Note that the TAP mean-field free energy~\cite{Morita1976,thouless1977solution} can be obtained via the expansion up to the second order of \eref{eq:GibbsFreeEnergy_alpha}~\cite{Plefka}.

\subsection{Adaptive Thouless--Anderson--Palmer equation}
\label{sec:AdaptiveTAPEquation_pairwiseMRF}

In this section, we show the derivation of the adaptive TAP equation for the quadratic MRF defined in \sref{sec:pairwiseMRF} via the conventional method: minimization of the adaptive TAP free energy.
There are several approaches for deriving the adaptive TAP free energy for the quadratic MRF~\cite{adaTAP1,adaTAP2,Csato,ATAPFE}.
Here, we focus on the approach based on the strategies proposed by Opper and Winther~\cite{adaTAP1,adaTAP2}.
The adaptive TAP free energy is defined as
\begin{align}
G_{\mathrm{adaTAP}}(\bm{m}, \bm{v}) := G_0(\bm{m},\bm{v}) + G_{1}^{\mathrm{GMRF}}(\bm{m},\bm{v}) - G_{0}^{\mathrm{GMRF}}(\bm{m},\bm{v}),
\label{eq:AdaptiveTAPFreeEnergy_definition}
\end{align}
where the first term on the right-hand side of \eref{eq:AdaptiveTAPFreeEnergy_definition} is \eref{eq:GibbsFreeEnergy_alpha} with $\alpha = 0$.
$G_{\alpha}^{\mathrm{GMRF}}(\bm{m},\bm{v})$ in \eref{eq:AdaptiveTAPFreeEnergy_definition} is \eref{eq:GibbsFreeEnergy_alpha}
when $\mathcal{X}=(-\infty, +\infty)$,
which corresponds to the Gaussian MRF~\cite{GMRF}.
Using Gaussian integration, we have
\begin{align}
&G_{\alpha}^{\mathrm{GMRF}}(\bm{m},\bm{v}) = - \sum_{i\in V}h_i m_i + \frac{1}{2}\sum_{i\in V}d_i v_i
+ \max_{\bm{\lambda}, \bm{\Lambda}}\left\{ \sum_{i\in V}\lambda_i m_i \right.\nonumber\\
&\left. - \frac{1}{2}\sum_{i\in V}\Lambda_i v_i- \frac{1}{2} \bm{\lambda}^{\mathrm{T}}\bm{S}_{\alpha}(\bm{\Lambda})^{-1}\bm{\lambda}
+ \frac{1}{2}\ln \det \bm{S}_{\alpha}(\bm{\Lambda}) \right\},
\label{eq:GaussianGibbsFreeEnergy_alpha}
\end{align}
where parameters $\bm{\lambda} = \{\lambda_i \mid i\in V\}$ and $\bm{\Lambda} = \{\Lambda_i\mid i\in V\}$ originate from the Lagrange multipliers
corresponding to the first and second constraints in \eref{eq:GibbsConstraints_Moments}, respectively.
Here, $\bm{S}_{\alpha}(\bm{\Lambda})$ is a symmetric matrix whose $ij$th element is defined by
\begin{align*}
 \big[\bm{S}_{\alpha}(\bm{\Lambda})\big]_{ij} := \left\{
 \begin{array}{ll}
 \Lambda_i & (i = j) \\
 -\alpha J_{ij} & \left((i,j)\in E\right)\\
 0 & (\mathrm{otherwise})
 \end{array}. \right.
\end{align*}
Executing the maximization with respect to $\bm{\lambda}$, \eref{eq:GaussianGibbsFreeEnergy_alpha} becomes
\begin{align}
&G_{\alpha}^{\mathrm{GMRF}}(\bm{m},\bm{v}) = - \sum_{i\in V}h_i m_i + \frac{1}{2}\sum_{i\in V}d_i v_i\nonumber\\
& +\max_{\bm{\Lambda}}\left\{ \frac{1}{2}\bm{m}^{\mathrm{T}}\bm{S}_{\alpha}(\bm{\Lambda})\bm{m} - \frac{1}{2}\sum_{i\in V}\Lambda_i v_i
+ \frac{1}{2}\ln \det\bm{S}_{\alpha}(\bm{\Lambda})\right\}.
\label{eq:GaussianGibbsFreeEnergy_alpha-type2}
\end{align}
From Eqs.~(\ref{eq:GibbsFreeEnergy_alpha}) and (\ref{eq:GaussianGibbsFreeEnergy_alpha-type2}),
\eref{eq:AdaptiveTAPFreeEnergy_definition} is obtained as
\begin{align}
&G_{\mathrm{adaTAP}}(\bm{m}, \bm{v}) = - \sum_{i\in V}h_i m_i + \frac{1}{2}\sum_{i\in V}d_i v_i + \sum_{i\in V} \hat{b}_i(0) m_i\nonumber\\
& - \sum_{i\in V} \hat{c}_i(0) v_i - \frac{1}{2} \sum_{i\in V}\ln(v_i - m_i^2) - \ln Z_0(\hat{\bm{b}}(0), \hat{\bm{c}}(0))\nonumber\\
& + \frac{1}{2}\max_{\bm{\Lambda}}\Big\{ \bm{m}^{\mathrm{T}}\bm{S}_{1}(\bm{\Lambda})\bm{m} - \sum_{i\in V}\Lambda_i v_i + \ln \det \bm{S}_{1}(\bm{\Lambda})\Big\}.
\label{eq:AdaptiveTAPFreeEnergy}
\end{align}

The adaptive TAP equation corresponds to the minimum condition of \eref{eq:AdaptiveTAPFreeEnergy} with respect to $\bm{m}$ and $\bm{v}$.
Hereinafter in this section, we denote the values of $\bm{m}$ and $\bm{v}$ at the minimum of \eref{eq:AdaptiveTAPFreeEnergy} by $\hat{\bm{m}}$ and $\hat{\bm{v}}$, respectively.
From the minimum conditions of \eref{eq:AdaptiveTAPFreeEnergy} with respect to $m_i$ and $v_i$, we obtain
\begin{align}
\hat{b}_i(0) &= h_i + \sum_{j \in \partial(i)}J_{ij}\hat{m}_j - \Big(\hat{\Lambda}_i + \frac{1}{\hat{v}_i - \hat{m}_i^2}\Big) \hat{m}_i
\label{eq:minimumCondition_aTAP_m}
\end{align}
and
\begin{align}
\hat{c}_i(0) &= d_i - \hat{\Lambda}_i - \frac{1}{\hat{v}_i - \hat{m}_i^2},
\label{eq:minimumCondition_aTAP_v}
\end{align}
respectively, where $\partial(i) := \{j \mid \{i,j\} \in E\}$ denotes the set of vertices that have a connection with $i$ and
$\hat{\bm{\Lambda}}$ denotes the solution to the maximum condition for $\bm{\Lambda}$ in \eref{eq:AdaptiveTAPFreeEnergy}:
\begin{align}
[\bm{S}_1(\hat{\bm{\Lambda}})^{-1}]_{ii} = \hat{v}_i - \hat{m}_i^2, \quad \forall i \in V.
\label{eq:maximumCondition_aTAP_Lambda}
\end{align}
Furthermore, from Eqs.~(\ref{eq:maximumCondition_m_alpha=0}) and (\ref{eq:maximumCondition_v_alpha=0}), we have
\begin{align}
\hat{m}_i &= \frac{\sum_{x_i \in \mathcal{X}} x_i \exp\big(\hat{b}_i(0) x_i - \hat{c}_i(0) x_i^2 / 2\big)}
{\sum_{x \in \mathcal{X}} \exp\big(\hat{b}_i(0) x - \hat{c}_i(0) x^2 / 2\big)},
\label{eq:adaptiveTAPEquation-m} \\
\hat{v}_i &= \frac{\sum_{x_i \in \mathcal{X}} x_i^2 \exp\big(\hat{b}_i(0) x_i - \hat{c}_i(0) x_i^2 / 2\big)}
{\sum_{x \in \mathcal{X}} \exp\big(\hat{b}_i(0) x - \hat{c}_i(0) x^2 / 2\big)}.
\label{eq:adaptiveTAPEquation-v}
\end{align}
Eqs. \eqref{eq:minimumCondition_aTAP_m}--\eqref{eq:adaptiveTAPEquation-v} represent the adaptive TAP equation.
We can obtain the approximate values of $\langle x_i \rangle$ and $\langle x_i^2 \rangle$ by solving the simultaneous equations with respect to $\hat{\bm{m}}$ and $\hat{\bm{v}}$, respectively.
However, the adaptive TAP equation includes matrix inversion (cf. \eref{eq:maximumCondition_aTAP_Lambda}).
This tends to obstruct the effective implementation of the adaptive TAP equation.

When $\mathcal{X} = \{-1,+1\}$ (i.e., when \eref{eq:pairwiseMRF} is an Ising model),
we have an alternative method of deriving the adaptive TAP equation
using I-SusP with the naive mean-field equation~\cite{I-SusP,jack2013meanfield}.
The adaptive TAP equation derived via I-SusP takes a message-passing type of formula, which does not explicitly include matrix inversion.
This simplifies the implementation of the adaptive TAP equation.

However, in quadratic MRFs, the equivalence of the adaptive TAP equation and the approximate equation obtained from the naive mean-field approximation with diagonal consistency has not been explicitly shown beyond an Ising model.
In the next section, we show that the naive mean-field approximation with diagonal consistency derives the approximate equation, which is equivalent to the adaptive TAP equation obtained in this section.

\section{Naive mean-field approximation with diagonal consistency for quadratic Markov random field}
\label{sec:I-SusP_pairwiseMRF}

In this section, we derive the approximate equation, which is the minimum condition of \eref{eq:NaiveMeanFieldFreeEnergy}, via naive mean-field approximation with diagonal consistency.
We show that it is equivalent to the adaptive TAP equation described in the previous section.

Let us consider the conventional SusP for the naive mean-field approximation.
SusP is a message-passing type of method for obtaining approximations of susceptibilities (or covariances) $\chi_{ij}^{\mathrm{exact}} := \langle x_i x_j \rangle - \langle x_i \rangle \langle x_j \rangle$.
From the linear response relation, we have $\chi_{ij}^{\mathrm{exact}} = \partial \langle x_i \rangle / \partial h_j$.
SusP uses its approximation, $\chi_{ij}^{\mathrm{exact}} \approx \chi_{ij}^{\mathrm{app}}= \partial m_i^{\mathrm{app}} / \partial h_j$,
where $m_i^{\mathrm{app}}$ is an approximation of $\langle x_i \rangle $ obtained using a method such as the naive mean-field approximation.
However, the susceptibilities obtained in this manner may not satisfy diagonal consistency.
This implies that the relations
\begin{align}
\chi_{ii}^{\mathrm{app}} = v_i^{\mathrm{app}}-(m_i^{\mathrm{app}})^2, \quad \forall i \in V,
\label{eq:diagonalConsistency}
\end{align}
may not hold, where $v_i^{\mathrm{app}}$ is an approximation of $\langle x_i^2 \rangle $ obtained by employing the same method as that used for obtaining $m_i^{\mathrm{app}}$.

In I-SusP, we incorporate the diagonal trick method into SusP to satisfy the diagonal consistency in \eref{eq:diagonalConsistency}~\cite{I-SusP,Yasuda2013,Yasuda2014}.
In this study, to derive approximate equations via the same computation as I-SusP with naive mean-field approximation, we extend the naive mean-field free energy in \eref{eq:NaiveMeanFieldFreeEnergy} as
\begin{align}
\tilde{G}_{\mathrm{naive}}(\bm{m},\bm{v}) := G_{\mathrm{naive}}(\bm{m},\bm{v}) - \frac{1}{2}\sum_{i \in V}\Lambda_i^{\dagger}\big(v_i - m_i^2 \big),
\label{eq:FreeEnergy_naive_I-SusP}
\end{align}
where $\bm{\Lambda}^{\dagger} := \{\Lambda_i^{\dagger} \mid i\in V\}$ are the auxiliary parameters that are determined to satisfy the diagonal consistency in \eref{eq:diagonalConsistency}.
For fixed $\bm{\Lambda}^{\dagger}$, we again denote the values of $\bm{m}$ and $\bm{v}$ at the minimum of \eref{eq:FreeEnergy_naive_I-SusP} by $\hat{\bm{m}}$ and $\hat{\bm{v}}$, respectively.
The minimum conditions of \eref{eq:FreeEnergy_naive_I-SusP} with respect to $m_i$ and $v_i$ lead to
\begin{align}
\hat{b}_i(0) &= h_i + \sum_{j \in \partial(i)}J_{ij}\hat{m}_j - \Lambda^{\dagger}_i \hat{m}_i,
\label{eq:minimumCondition_I-SusP_m}
\end{align}
and
\begin{align}
\hat{c}_i(0) &= d_i -\Lambda^{\dagger}_i,
\label{eq:minimumCondition_I-SusP_v}
\end{align}
respectively. The relations between $\{\hat{m}_i, \hat{v}_i\}$ and $\{\hat{b}_i(0), \hat{c}_i(0)\}$ are already given in Eqs.~\eqref{eq:adaptiveTAPEquation-m} and \eqref{eq:adaptiveTAPEquation-v}.
Approximate susceptibilities are obtained via the linear response relation, $\chi_{ij}:= \partial \hat{m}_i / \partial h_j$.
Therefore, from Eqs.~\eqref{eq:adaptiveTAPEquation-m} and \eqref{eq:minimumCondition_I-SusP_m}, we obtain the simultaneous equations for the susceptibilities as
\begin{align}
\chi_{ij} = \frac{\hat{v}_i - \hat{m}_i^2}{1 + \Lambda_i^{\dagger}\big(\hat{v}_i - \hat{m}_i^2\big)}
\Big(\delta_{ij} + \sum_{k\in \partial(i)}J_{ik}\chi_{kj}\Big),
\label{eq:susceptibility_I-SusP}
\end{align}
where $\delta_{ij}$ is the Kronecker delta.
As mentioned earlier, $\bm{\Lambda}^{\dagger}$ should be determined to satisfy the diagonal consistency, $\chi_{ii} = \hat{v}_i - \hat{m}_i^2$.
Therefore, they are determined by
\begin{align}
\Lambda_i^{\dagger} = \frac{1}{\hat{v}_i - \hat{m}_i^2}\sum_{k \in \partial (i)}J_{ik}\chi_{ki}.
\label{eq:diagonalConsistency_I-SusP}
\end{align}
This is obtained from \eref{eq:susceptibility_I-SusP} and $\chi_{ii} = \hat{v}_i - \hat{m}_i^2$.
Solving Eqs.~\eqref{eq:adaptiveTAPEquation-m}, \eqref{eq:adaptiveTAPEquation-v}, and \eqref{eq:minimumCondition_I-SusP_m}--\eqref{eq:diagonalConsistency_I-SusP}
with respect to $\hat{\bm{m}}$, $\hat{\bm{v}}$, and $\bm{\chi}$ provides the approximations for the first-order moments, second-order moments, and susceptibilities, respectively.

The equivalence of the solutions to the adaptive TAP equation (Eqs.~\eqref{eq:minimumCondition_aTAP_m}--\eqref{eq:adaptiveTAPEquation-v})
and the approximate equation obtained from the naive mean-field approximation with diagonal consistency (Eqs.~\eqref{eq:adaptiveTAPEquation-m}, \eqref{eq:adaptiveTAPEquation-v}, and \eqref{eq:minimumCondition_I-SusP_m}--\eqref{eq:diagonalConsistency_I-SusP}) can be easily verified.
By considering $\hat{\Lambda}_i = \Lambda_i^{\dagger} + 1 / (\hat{v}_i - \hat{m}_i^2) $, Eqs.~\eqref{eq:minimumCondition_I-SusP_m} and \eqref{eq:minimumCondition_I-SusP_v} become Eqs.~\eqref{eq:minimumCondition_aTAP_m} and \eqref{eq:minimumCondition_aTAP_v}.
Furthermore, from \eref{eq:susceptibility_I-SusP}, we obtain
\begin{align*}
\delta_{ij} = \sum_{k\in V} \left( \delta_{ik}\hat{\Lambda}_i - J_{ik} \right) \chi_{kj}
=\sum_{k\in V} \big[\bm{S}_{1}(\hat{\bm{\Lambda}}) \big]_{i,k}\chi_{kj}.
\end{align*}
This implies that matrix $\bm{\chi}$ is equivalent to the inverse of $\bm{S}_{1}(\hat{\bm{\Lambda}})$.
On the contrary, the diagonal susceptibilities obtained from the naive mean-field approximation with diagonal consistency satisfy $\chi_{ii} = \hat{v}_i - \hat{m}_i^2$.
Therefore, $\chi_{ii} =[\bm{S}_{1}(\hat{\bm{\Lambda}})^{-1}]_{ii}= \hat{v}_i - \hat{m}_i^2$ (\eref{eq:maximumCondition_aTAP_Lambda}) is ensured.
Based on the above, we found that the solutions to the adaptive TAP equation and the approximate equation obtained from naive mean-field approximation with diagonal consistency are generally equivalent in quadratic MRFs.
This result supports the validity of our prediction about the equivalence of the adaptive TAP approximation and I-SusP with the naive mean-field approximation.
Even though the equivalence has not been rigorously proven yet, we move to the following arguments by accepting it as the ``ansatz''.


The advantage of the naive mean-field approximation with diagonal consistency compared with the conventional adaptive TAP approximation is that it is considerably easier to apply this method to models beyond quadratic MRFs, such as higher-order MRFs.
In the typical approaches to the adaptive TAP equation~\cite{adaTAP1,adaTAP2,Csato,ATAPFE}, it is essential for the energy function to be quadratic, implying that applying such approaches to higher-order MRFs is not straightforward.
In contrast, the naive mean-field approximation with diagonal consistency can be applied to models whose naive mean-field approximation can be explicitly described.
This implies that we can consider an adaptive-TAP-like approximate equation in models beyond quadratic MRFs, such as higher-order MRFs, via the naive mean-field approximation with diagonal consistency.

On the other hand, the naive mean-field approximation with diagonal consistency has no particular advantage compared with the conventional adaptive TAP approximation in terms of computational complexity.
If the MRF is a fully-connected model, the naive mean-field approximation with diagonal consistency costs $O(n^3)$ similar to the conventional adaptive TAP approach.
If the MRF is not a fully-connected model, for example, it is defined on a sparse graph such as a square-lattice (i.e., the expectation of the degree of the graph is at most $O(1)$), the naive mean-field approximation with diagonal consistency costs $O(n^2)$.

%

\section{Adaptive Thouless--Anderson--Palmer equation for higher-order Markov random field}
\label{sec:adaptiveTAPApproach_higher-order}

\subsection{Higher-order Boltzmann machine\\and its Gibbs free energy}
\label{sec:higherOrderMRF}

We consider distinct subgraphs, $\mu \subseteq V$, in $\mathcal{G}(V,E)$ and denote the family of all the subgraphs by $\mathcal{C}$.
Let us consider an MRF with higher-order interactions whose energy function is described as
\begin{align}
H^{\star}(\bm{x}) := - \sum_{i\in V} h_i x_i + \frac{1}{2}\sum_{i\in V} d_i x_i^2 - \sum_{\mu \in \mathcal{C}}J_{\mu}\prod_{i \in \mu}x_i,
\label{eq:energy_higher-order}
\end{align}
where $J_{\mu}$ is the interaction weight among the vertices contained in $\mu$.
When all subgraphs in $\mathcal{C}$ are connected pairs in $\mathcal{G}(V,E)$, \eref{eq:energy_higher-order} is reduced to \eref{eq:energy}.
This MRF is known as the higher-order Boltzmann machine~\cite{HBM1987}.

In a manner similar to \sref{sec:GibbsFreeEnergy}, we can derive the GFE and naive mean-field free energy for this higher-order Boltzmann machine.
The GFE with auxiliary parameter $\alpha \in [0,1]$ for adjusting the effect of the interaction is expressed as
\begin{align}
&G_{\alpha}^{\star}(\bm{m}, \bm{v}) = - \sum_{i\in V} h_i m_i + \frac{1}{2} \sum_{i\in V} d_i v_i
+ \max_{\bm{b},\bm{c}}\Big\{ \sum_{i\in V} b_i m_i \nonumber\\
& - \frac{1}{2}\sum_{i\in V} c_i v_i - \ln \sum_{\bm{x} \in \mathcal{X}^n} \exp \Big( \sum_{i\in V}b_i x_i - \frac{1}{2}\sum_{i\in V}c_i x_i^2 \nonumber\\
& + \alpha\sum_{\mu \in \mathcal{C}}J_{\mu}\prod_{i \in \mu}x_i\Big) \Big\}.
\label{eq:GibbsFreeEnergy_higher-order_alpha}
\end{align}
The Plefka expansion for \eref{eq:GibbsFreeEnergy_higher-order_alpha} provides the naive mean-field free energy as
\begin{align}
&G_{\mathrm{naive}}^{\star}(\bm{m},\bm{v}):= - \sum_{i\in V}h_i m_i + \frac{1}{2}\sum_{i\in V}d_i v_i + \sum_{i\in V} \hat{b}_i(0)m_i\nonumber\\
& - \frac{1}{2}\sum_{i\in V}\hat{c}_i(0) v_i - \ln Z_0(\hat{\bm{b}}(0), \hat{\bm{c}}(0))
- \sum_{\mu \in \mathcal{C}}J_{\mu}\prod_{i \in \mu}m_i,
\label{eq:NaiveMeanFieldFreeEnergy_higher-order}
\end{align}
where $Z_0(\hat{\bm{b}}(0), \hat{\bm{c}}(0))$ is already defined in \eref{eq:Z_alpha=0}.
Parameters $\hat{\bm{b}}(0)$ and $\hat{\bm{c}}(0)$ satisfy Eqs.~\eqref{eq:maximumCondition_m_alpha=0} and \eqref{eq:maximumCondition_v_alpha=0}.
The naive mean-field equation is obtained from the minimum conditions of \eref{eq:NaiveMeanFieldFreeEnergy_higher-order} with respect to $\bm{m}$ and $\bm{v}$.

\subsection{Adaptive Thouless--Anderson--Palmer equation\\for higher-order Boltzmann machine}
\label{sec:adaptiveTAPEquation_higher-order}

In a manner similar to \sref{sec:I-SusP_pairwiseMRF}, we derive the adaptive-TAP-like equation for the higher-order Boltzmann machine via naive mean-field approximation with diagonal consistency.
Similar to \eref{eq:FreeEnergy_naive_I-SusP}, we extend the naive mean-field free energy in \eref{eq:NaiveMeanFieldFreeEnergy_higher-order} by installing the diagonal trick term:
\begin{align}
\tilde{G}_{\mathrm{naive}}^{\star}(\bm{m},\bm{v}) := G_{\mathrm{naive}}^{\star}(\bm{m},\bm{v}) - \frac{1}{2}\sum_{i \in V}\Lambda_i^{\ddagger}\big(v_i - m_i^2 \big).
\label{eq:FreeEnergy_naive_I-SusP_higher-order}
\end{align}
For fixed $\bm{\Lambda}^{\ddagger}:= \{\Lambda_i^{\ddagger} \mid i\in V\}$, we again denote the values of $\bm{m}$ and $\bm{v}$ at the minimum of \eref{eq:FreeEnergy_naive_I-SusP_higher-order}
by $\hat{\bm{m}}$ and $\hat{\bm{v}}$, respectively.
The minimum conditions of \eref{eq:FreeEnergy_naive_I-SusP_higher-order} with respect to $m_i$ and $v_i$ lead to
\begin{align}
\hat{b}_i(0) &= h_i + \sum_{\mu \in \mathcal{C}(i)} J_{\mu} \prod_{j \in \mu \setminus \{i\}}\hat{m}_j - \Lambda^{\ddagger}_i \hat{m}_i,
\label{eq:minimumCondition_I-SusP_m_higher-order}
\end{align}
and
\begin{align}
\hat{c}_i(0) &= d_i -\Lambda^{\ddagger}_i,
\label{eq:minimumCondition_I-SusP_v_higher-order}
\end{align}
respectively, where $\mathcal{C}(i)\subseteq \mathcal{C}$ is the family of the subgraphs containing $i$.
The relations between $\{\hat{m}_i ,\hat{v}_i\}$ and $\{\hat{b}_i(0) , \hat{c}_i(0)\}$ are already given in Eqs.~\eqref{eq:adaptiveTAPEquation-m} and \eqref{eq:adaptiveTAPEquation-v}.
From Eqs.~\eqref{eq:adaptiveTAPEquation-m} and \eqref{eq:minimumCondition_I-SusP_m_higher-order}, the linear response relation, $\chi_{ij}= \partial \hat{m}_i / \partial h_j$, is obtained as
\begin{align}
\chi_{ij} &= \frac{\hat{v}_i - \hat{m}_i^2}{1 + \Lambda_i^{\ddagger}\big(\hat{v}_i - \hat{m}_i^2\big)}\nonumber\\
&\quad \times
\Big(\delta_{ij} + \sum_{\mu \in \mathcal{C}(i)} J_{\mu}\sum_{k \in \mu \setminus \{i\}}\chi_{kj}
\prod_{l \in \mu \setminus \{i, k\}}\hat{m}_l \Big).
\label{eq:susceptibility_I-SusP_higher-order}
\end{align}
Finally, combining the diagonal consistency, $\chi_{ii} = \hat{v}_i - \hat{m}_i^2$, with \eref{eq:susceptibility_I-SusP_higher-order} provides the equations for determining $\bm{\Lambda}_i^{\ddagger}$ as
\begin{align}
\Lambda_i^{\ddagger} = \frac{1}{\hat{v}_i - \hat{m}_i^2}\sum_{\mu \in \mathcal{C}(i)} J_{\mu}\sum_{k \in \mu \setminus \{i\}}\chi_{ki}
\prod_{l \in \mu \setminus \{i, k\}}\hat{m}_l.
\label{eq:diagonalConsistency_I-SusP_higher-order}
\end{align}
Solving Eqs.~\eqref{eq:adaptiveTAPEquation-m}, \eqref{eq:adaptiveTAPEquation-v}, and \eqref{eq:minimumCondition_I-SusP_m_higher-order}--\eqref{eq:diagonalConsistency_I-SusP_higher-order} with respect to $\hat{\bm{m}}$, $\hat{\bm{v}}$, and $\bm{\chi}$ simultaneously provides the approximations for the first-order moments, second-order moments, and susceptibilities, respectively, for the higher-order Boltzmann machine.
Note that Eqs.~\eqref{eq:adaptiveTAPEquation-m}, \eqref{eq:adaptiveTAPEquation-v}, \eqref{eq:minimumCondition_I-SusP_m_higher-order} and \eqref{eq:minimumCondition_I-SusP_v_higher-order} incorporate the effect of diagonal consistency and contribute to the inference.
In Reference~\cite{leisink2000learning}, the inference and learning of the higher-order Boltzmann machine have been generalized.
In their work, while the learning is constrained by the ``diagonal couplings'' corresponding to the diagonal consistency, the inference is done by usual mean-field equations that do not include such constraints.
In this respect, the proposing method in this section and the method in Reference~\cite{leisink2000learning} are fundamentally different.
(A detailed derivation of the learning using the approximate equations in this section is omitted here.)


For $p$-spin Sherrington--Kirkpatrick model~\cite{derrida1980,derrida1981}, the naive mean-field approximation with diagonal consistency yields the TAP equation presented in Reference~\cite{rieger1992number}.
The details are shown in \aref{app:p_spin_TAP}.

\subsection{Numerical experiments}
\label{sec:experiments}

In this section, we demonstrate the performance of the naive mean-field approximation with diagonal consistency presented in \sref{sec:adaptiveTAPEquation_higher-order}.
In the experiments, we consider a higher-order Boltzmann machine whose energy function is
\begin{align}
H^{\star}(\bm{x}) &= - \sum_{i\in V} h_i x_i + \frac{d}{2}\sum_{i\in V} x_i^2 - \sum_{\{i, j\}\in \mathcal{C}_2} J_{ij} x_i x_j \nonumber\\
&\quad - J_3\sum_{\{i, j, k\}\in \mathcal{C}_3} x_i x_j x_k,
\label{eq:energy_higher-order_for_experiment}
\end{align}
where $\mathcal{C}_2 := \{ \{i,j\} \mid i \in V, j \in V, i<j\}$ is the family of all distinct pairs
and $\mathcal{C}_3 := \{ \{i,j, k\} \mid i \in V, j \in V, k \in V, i<j < k\}$ is the family of all distinct triplets.
\eref{eq:energy_higher-order_for_experiment} is a special case of \eref{eq:energy_higher-order}.
When $J_3 = 0$, \eref{eq:energy_higher-order_for_experiment} is reduced to the quadratic energy shown in \eref{eq:energy}.
In the experiments described below, we set $n=10$ and $J_3 = 0.001$.
Parameters $\{h_i\}$ and $\{J_{ij}\}$ were independently drawn from Gaussian distributions $\mathcal{N}(0, 0.1^2)$ and $\mathcal{N}(0, \sigma^2/\sqrt{n})$, respectively, where $\mathcal{N}(\mu, \sigma^2)$ is the Gaussian distribution with mean $\mu$ and variance $\sigma^2$.
As the number of variables is not large in this setting, we can compute exact expectations and compare them with approximate expectations.
We used mean squared errors (MSEs) defined by
\begin{align*}
 \mathrm{MSE}_{m} &:= \frac{1}{n}\sum_{i\in V}\left(\langle x_i \rangle - m_i^{\mathrm{app}}\right)^2,\\
 \mathrm{MSE}_{v} &:= \frac{1}{n}\sum_{i\in V}\left(\left\langle x_i^2 \right\rangle - v_i^{\mathrm{app}}\right)^2,
\end{align*}
as the performance measure.
We compared the solutions to the naive mean-field approximation with diagonal consistency presented in \sref{sec:adaptiveTAPEquation_higher-order} and the simple naive mean-field approximation in terms of the MSEs.
The solution to the simple naive mean-field approximation for \eref{eq:energy_higher-order_for_experiment} is obtained from the minimum conditions of \eref{eq:NaiveMeanFieldFreeEnergy_higher-order} with respect to $\bm{m}$ and $\bm{v}$.

\begin{figure}[t]
\centering
\includegraphics[width=5.5cm]{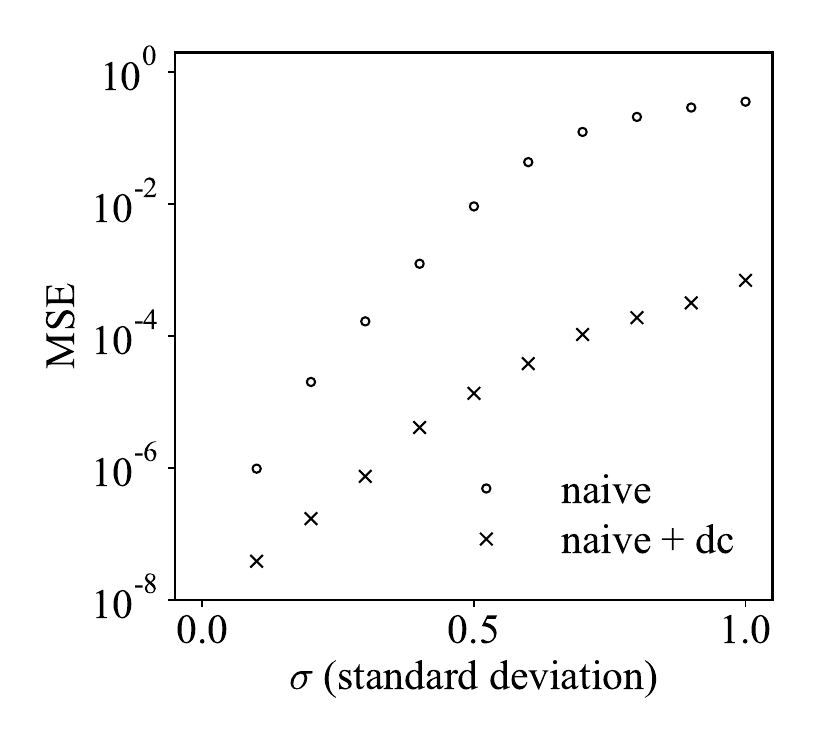}
\caption{Plot of $\mathrm{MSE}_m$ against $\sigma$ when $\mathcal{X} = \{-1, +1\}$.
The points labeled as ``naive'' and ``naive + dc'' are the results obtained by the simple naive mean-field approximation and the naive mean-field approximation with diagonal consistency, respectively.
Each point in the plot denotes the average value over 1000 trials.}
\label{fig:experiment1}
\end{figure}

Figure \ref{fig:experiment1} shows the result of $\mathrm{MSE}_m$ against $\sigma$ when $\mathcal{X} = \{-1, +1\}$.
In this case, as $x_i^2$ is always one, $v_i$ is also always one in both methods.
Hence, $\mathrm{MSE}_v$ is always zero.
It is noteworthy that the value of $d$ is unrelated to the result because the second term in \eref{eq:energy_higher-order_for_experiment} is constant.
\begin{figure}[t]
\centering
\vspace{10pt}
\includegraphics[width=8.6cm]{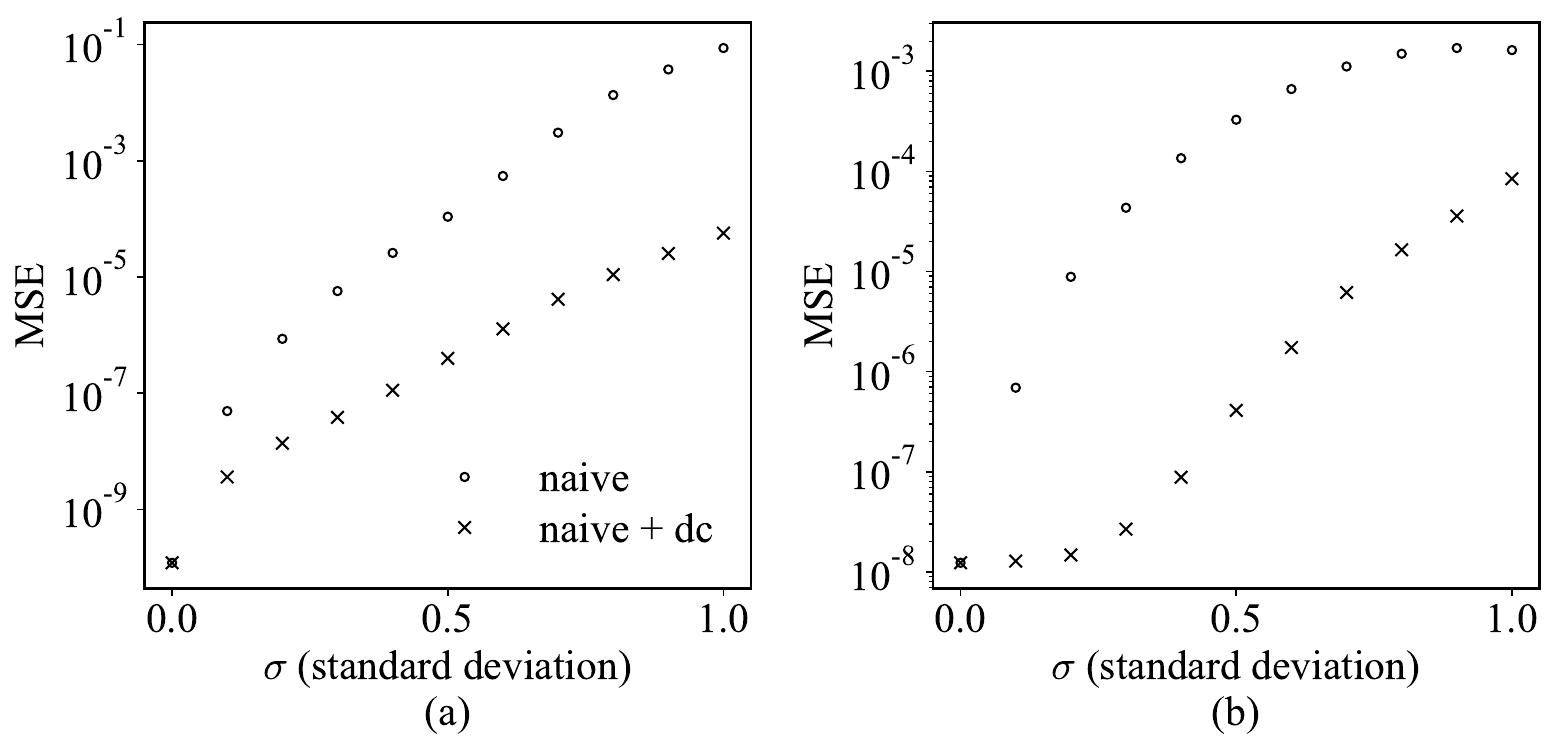}
\caption{
Plots of (a) $\mathrm{MSE}_m$ and (b) $\mathrm{MSE}_v$ against $\sigma$ when $\mathcal{X} = \{-1, 0,+1\}$.
The points labeled as ``naive'' and ``naive + dc'' are the results obtained by the simple naive mean-field approximation and the naive mean-field approximation with diagonal consistency, respectively.
Each point in the plot denotes the average value over 1000 trials.
}
\label{fig:experiment2}
\end{figure}
Figure \ref{fig:experiment2} shows the result of (a) $\mathrm{MSE}_m$ and (b) $\mathrm{MSE}_v$ against $\sigma$ when $\mathcal{X} = \{-1, 0,+1\}$.
In this experiment, we set $d = 0.01$.
The naive mean-field approximation with diagonal consistency outperforms the simple naive mean-field approximation in both experiments, as we expected.

\section{Discussion and conclusion}
\label{sec:conclusion}

We have formulated the adaptive-TAP-like approximate equation for higher-order Boltzmann machines via the naive mean-field approximation with diagonal consistency.
In the numerical experiments, we have observed that the expectations of the variables obtained using the adaptive-TAP-like equations are more accurate than those obtained using the simple naive mean-field approximation.
It is noteworthy that a method almost the same as I-SusP was independently proposed around the same time by Raymond and Ricci-Tersenghi~\cite{jack2013meanfield}.
While I-SusP considers only diagonal consistency, their method additionally involves constraints with regard to off-diagonal consistency~\cite{jack2017improving}.
Our approach can be extended by employing such advanced constraints.

In addition, in \sref{sec:adaptiveTAPApproach_higher-order}, only models whose Hamiltonian does not include higher-order terms, such as $x_i^2 x_j$ and $x_i^3$, are considered.
Such higher-order terms should also be considered if the variables have discrete or continuous values.
Because this generalization requires complicated formulations other than the procedure shown in \sref{sec:I-SusP_pairwiseMRF} and \sref{sec:adaptiveTAPApproach_higher-order}, details of the generalization are omitted in this paper.
This task will be addressed soon.

In this paper, we have reported the results of numerical performance evaluation for direct problems (inference).
The adaptive TAP equation or other mean-field approaches that use the linear response relation are known to be effective against the inverse problem (learning)~\cite{leisink2000learning,huang2013adaptive,kiwata2014estimation,gabrie2015training}.
Application of the adaptive-TAP-like equation to the inverse problem and its performance evaluation will be addressed in our future tasks.

Another challenge to address is the theoretical verification of the performance of the naive mean-field approximation with diagonal consistency.
The adaptive TAP equation or the adaptive-TAP-like equation frequently converges more slowly compared to the simple naive mean-field equation or the simple TAP equation or fails to converge depending on the settings of problems.
In this study, the accuracy of the approximation has been clarified only from the experimental aspect.
Several mean-field-based algorithms whose performance has been guaranteed theoretically have been proposed in previous studies~\cite{yuille2002cccp,chako2016meanfield}.
Overcoming this challenge will improve our understanding of I-SusP, or the naive mean-field method proposed in this study, and the adaptive TAP approximation.

\begin{acknowledgments}
The authors would like to thank Shun Kataoka and Yuya Seki for their insightful comments and suggestions.

The authors were supported by CREST, Japan Science and Technology Agency Grant (No. JPMJCR1402).
One of the authors (C. T.) was partially supported by a Grant-in-Aid for JSPS Fellows from the Japan Society for the Promotion of Science Grant (No. JP17J03081).
One of the authors (M. Y.) was partially supported by a Grant-in-Aid for Scientific Research from the Japan Society for the Promotion of Science Grant (Nos. 15K00330, 15H03699, 18K11459 and 18H03303).
One of the authors (K. T.) was partially supported by a Grant-in-Aid for Scientific Research from the Japan Society for the Promotion of Science Grant (No. 18H03303).
\end{acknowledgments}

\onecolumngrid
\appendix

\section{TAP equation of $p$-spin Ising Sherrington-Kirkpatrick model}
\label{app:p_spin_TAP}

For $p$-spin Ising Sherrington-Kirkpatrick (SK) model~\cite{derrida1980,derrida1981}, the approximate equation derived by naive mean-field approximation with diagonal consistency reproduces the TAP equation.
Following the notation in \eref{eq:energy_higher-order}, the Hamiltonian of a $p$-spin Ising SK model can be written as
\begin{align}
  H_p(\bm{x}) = - \sum_{i\in V} h_i x_i - \sum_{\mu \in \mathcal{C}} J_{\mu} \prod_{i \in \mu} x_{i},
  \label{eq:Hamiltonian_p_spin}
\end{align}
where $x_i \in \{-1, +1\}$ are $N$ Ising spins, $\mu = \{i_1, i_2, \cdots, i_p\}$ and $\mathcal{C} = \{\mu \mid i_1 \in V, i_2 \in V, \cdots, i_p \in V, \; i_1 < i_2 < \cdots < i_p\}$.
The couplings $J_{\mu}$ are the quenched random variables distributed according to a Gaussian distribution
\begin{align}
  P(J_{\mu}) = \frac{1}{\sqrt{\pi} \tilde{J}} \exp \left(- \frac{J_{\mu}^2}{\tilde{J}^2} \right), \quad \tilde{J}^2 = \frac{J^2 p!}{N^{p-1}}.
  \label{eq:gaussian_assumption_p_spin}
\end{align}
The $p$-spin Ising SK model is defined as
\begin{align}
  P_{p}(\bm{x}) = \frac{1}{Z_p} \exp (- H_p(\bm{x})).
  \label{eq:p_spin_Ising_SK}
\end{align}

The approximate equations derived by naive mean-field approximation with diagonal consistency are Eqs.~\eqref{eq:adaptiveTAPEquation-m}, \eqref{eq:minimumCondition_I-SusP_m_higher-order}, \eqref{eq:susceptibility_I-SusP_higher-order} and \eqref{eq:diagonalConsistency_I-SusP_higher-order}, where $\hat{v}_i = 1$.
We can reproduce the TAP equation of the $p$-spin Ising SK model in \eref{eq:p_spin_Ising_SK} by using the expansion with respect to the couplings.
We introduce the auxiliary parameter $\alpha \in [0, 1]$ that describes the strength of the interaction to \eref{eq:susceptibility_I-SusP_higher-order}, as
\begin{align}
  \chi_{ij}(\alpha) &:= \frac{1 - \hat{m}_i^2}{1 + \Lambda_i^{\ddagger}(\alpha)(1 - \hat{m}_i^2)} \left( \delta_{ij} + \alpha \sum_{\mu \in \mathcal{C}(i)} J_{\mu} \sum_{k \in \mu \setminus \{i\}} \chi_{kj}(\alpha) \prod_{l \in \mu \setminus \{i,k\}}\hat{m}_l \right),
  \label{eq:chi_alpha}
\end{align}
and to \eref{eq:diagonalConsistency_I-SusP_higher-order}, as
\begin{align}
  \Lambda_i^{\ddagger}(\alpha) := \frac{\alpha}{1 - \hat{m}_i^2}\sum_{\mu \in \mathcal{C}(i)} J_{\mu} \sum_{k \in \mu \setminus \{i\}} \chi_{ki}(\alpha) \prod_{l \in \mu \setminus \{i, k\}} \hat{m}_l,
  \label{eq:Lambda_alpha}
\end{align}
respectively.
From the Taylor expansion of Eqs.~\eqref{eq:chi_alpha} and \eqref{eq:Lambda_alpha} with respect to $\alpha$ (instead of the direct expansion with respect to the couplings), we obtain
\begin{align}
  \Lambda_{i}^{\ddagger}(\alpha) &= \alpha^2 \sum_{k_2} (1 - \hat{m}_{k_2}^2) \left( \sum_{\mu \in \mathcal{C}(i, k_2)} J_{\mu} \prod_{l \in \mu \setminus \{i, k_2\}}\hat{m}_l \right)^2 + \alpha^3 \sum_{k_2}\sum_{k_3} (1 - \hat{m}_{k_2}^2)(1 - \hat{m}_{k_3}^2) \left( \sum_{\mu \in \mathcal{C}(i, k_2, k_3)} J_{\mu} \prod_{l \in \mu \setminus \{i, k_2, k_3\}}\hat{m}_l \right)^3 \nonumber\\
  &\quad - \alpha^4 (1 - \hat{m}_i^2) \sum_{k_2} (1 - \hat{m}_{k_2}^2)^2 \left( \sum_{\mu\in\mathcal{C}(i, k_2)} J_{\mu} \prod_{l\in \mu\setminus\{i, k_2\}} \hat{m}_l \right)^4\nonumber\\
  &\quad + \alpha^4 \sum_{k_2} \sum_{k_3} \sum_{k_4} (1 - \hat{m}_{k_2}^2) (1 - \hat{m}_{k_3}^2) (1 - \hat{m}_{k_4}^2) \left( \sum_{\mu\in\mathcal{C}(i, k_2, k_3, k_4)} J_{\mu} \prod_{l\in \mu\setminus\{i, k_2, k_3, k_4\}} \hat{m}_l \right)^4 + O(\alpha^5),
  \label{eq:Lambda_alpha_}
\end{align}
based on Reference~\cite{Yasuda2007}.
Here, for distinct indices $i, k_2, \cdots, k_{\gamma}, \; \mathcal{C}(i, k_2, \cdots, k_{\gamma}) \subseteq \mathcal{C}$ is defined as the family of the subgraphs containing $i, k_2, \cdots, k_{\gamma}$, i.e., $\mathcal{C}(i, k_2, \cdots, k_{\gamma}) = \{ \mu \mid \{i, k_2, \cdots, k_r\} \subseteq \mu, \; \mu \in \mathcal{C} \}$, for $p \geq \gamma$.
For $p < \gamma$, we define $\mathcal{C}(i, k_2, \cdots, k_{\gamma}) = \emptyset$.
Therefore, $|\mathcal{C}(i, k_2, \cdots, k_{\gamma})| = O(N^{p-\gamma})$.
$\sum_{\mu \in \mathcal{C}(i, k_2, \cdots, k_{\gamma})}$ is $O(N^{p - \gamma})$ sums, where $\gamma \geq 2$ is the exponent of $\alpha$.
The sums $\sum_{\mu \in \mathcal{C}(i, k_2, \cdots, k_{\gamma})} J_{\mu} \prod_{l\in \mu\setminus\{i, k_2, \cdots, k_{\gamma}\}} \hat{m}_l$ can be regarded as a sum over $O(N^{p-\gamma})$ independent random variables with variance $O(N^{-(p-1)})$, i.e.,
are the Gaussian random variables with mean zero and variance $O(N^{-(\gamma - 1)})$ from the central limit theorem.
The $\gamma$-th power of the sums can be regarded as the independent random variables with mean $\mu_{\gamma} = 0$ (if $\gamma$ is odd) or $\mu_{\gamma} = O(N^{-\gamma(\gamma - 1)/2})$ (if $\gamma$ is even) and variance $\sigma_{\gamma}^2 = O(N^{-\gamma(\gamma - 1)})$.
In \eref{eq:Lambda_alpha_}, $O(\alpha^2)$, $O(\alpha^3)$, and the second term of $O(\alpha^4)$ consist of $\sum_{k_2} \cdots \sum_{k_{\gamma}}(\sum_{\mu \in \mathcal{C}(i, k_2, \cdots, k_{\gamma})} J_{\mu} \prod_{l\in \mu\setminus\{i, k_2, \cdots, k_{\gamma}\}} \hat{m}_l)^{\gamma}$, which are the Gaussian random variables with mean $O(N^{\gamma - 1}\mu_{\gamma})$ and variance $O(N^{\gamma -1}\sigma_{\gamma}^2)$.
In the first term of $O(\alpha^4)$, $\sum_{k_2} (\sum_{\mu \in \mathcal{C}(i, k_2)} J_{\mu} \prod_{l\in \mu\setminus\{i, k_2\}} \hat{m}_l)^{4}$ is the Gaussian random variable with mean $O(N \mu_{4})$ and variance $O(N \sigma_{4}^2)$.
Thus, the computational order of the term of $O(\alpha^2)$ is evaluated as $O(1)$.
On the other hand, the order of the terms of $O(\alpha^3)$ and $O(\alpha^4)$ are given by $O(N^{-1})$ and $O(N^{-2})$, respectively.
As $\gamma$ increases, the order of the higher-order terms becomes smaller with respect to $N$.
From this, higher-order terms $O(\alpha^{\gamma})\;(\gamma \geq 3)$ can be negligible in the $N\to\infty$ limit.
By setting $\alpha = 1$, we obtain
\begin{align}
  \Lambda_{i}^{\ddagger} = \sum_{k} (1 - \hat{m}_k^2) \left( \sum_{\mu \in \mathcal{C}(i, k)} J_{\mu} \prod_{l \in \mu \setminus \{i, k\}}\hat{m}_l \right)^2 + O(N^{-1}).
  \label{eq:onsager_p_spin_}
\end{align}
Equations \eqref{eq:adaptiveTAPEquation-m} and \eqref{eq:minimumCondition_I-SusP_m_higher-order} with \eref{eq:onsager_p_spin_} correspond to the TAP equation in Reference~\cite{rieger1992number}.

\vspace{7em}
\twocolumngrid
\bibliographystyle{apsrev4-1}
\bibliography{references}

\end{document}